\documentclass[prb,twocolumn,final,showpacs]{revtex4}

\usepackage{amsmath,amsfonts,amssymb,bm}
\usepackage{dcolumn}
\usepackage[final]{graphicx}
\usepackage{bm}

\begin{document}

\bibliographystyle{apsrev}	

\title{Influence of surface roughness on the optical properties of plasmonic nanoparticles}

\author{Andreas Tr\"ugler}
\author{Jean-Claude Tinguely}
\author{Joachim~R.~Krenn}
\author{Andreas Hohenau}
\author{Ulrich Hohenester}
\email{ulrich.hohenester@uni-graz.at}

\affiliation{Institut f\"ur Physik,
  Karl--Franzens--Universit\"at Graz, Universit\"atsplatz 5,
  8010 Graz, Austria}

\date{\today}

\begin{abstract}
For plasmonic nanoparticles, we investigate the influence of surface
roughness inherent to top-down fabrication on the optical properties,
and find that it has a surprisingly small influence on the position and
width of the plasmon peaks.  Our experimental observation is supported
by simulations based on the boundary element method approach.  Using a
perturbation approach, suitable for metallic nanoparticles with a
moderate degree of surface roughness, we demonstrate that the reason for
this lies in \textit{motional narrowing} where the plasmon averages over
the random height fluctuations.  Surface roughness in large arrays of
identical nanoparticles, such as encountered in the context of
metamaterials, is thus expected to not constitute a major roadblock.
\end{abstract}



\maketitle


Plasmonics bridges the gap between the micrometer length scale of light and the
length scale of nanostructures.~\cite{maier:07}
This is achieved by binding light to coherent charge density
oscillations of metallic nanostructures, so-called surface plasmons, 
which allow to focus electromagnetic
radiation down to spots with spatial dimensions of the order of a few
nanometers.~\cite{Schuller:NatureMat:2010}  Coupling of quantum
emitters, such as quantum dots or molecules, with plasmonic
nanostructures can strongly modify their excitation and emission
properties, observable in fluorescence~\cite{anger:06} or surface
enhanced Raman scattering,~\cite{Kneipp:Book} and
offers a unique means for tailoring light-matter interaction at the
nanoscale.  This has found widespread applications ranging from
(bio)sensors~\cite{anker:08} and solar cells~\cite{ferry:08} to
optical and quantum communication technology.~\cite{Chang:PhysRevLett:06} Plasmonic nanoparticles
are also at the heart of the emerging fields of metamaterials and
optical cloaking.~\cite{pendry:06}

Huge advances in fabrication techniques over the last years allow
nowadays to fabricate plasmonic nanostructures with well understood and
predictable properties.  Nevertheless, practically all metallic
nanoparticles suffer from size inhomogeneities and nanoscale surface
roughness,~\cite{Marzan:CPPC:2009, Chen:NanoLetters:09} which results in
deviations of the plasmonic properties from those of idealized
nanoparticles.~\cite{Barnes:JOptA:09}  Particularly top-down approaches
for nanoparticle fabrication often involve vacuum deposition of the
metal structures, which leads to polycrystalline particles with an
apparent surface roughness.~\cite{Chen:NanoLetters:09}  Despite the
important role of surface roughness,  there is still little
understanding about the impact of such imperfection on the optical
properties. Recent publications report on the control of nanoscale
roughness and its  strong effect on the nanoparticles far- and near-field 
optical  properties.~\cite{Chen:NanoLetters:09, Marzan:CPPC:2009,
Hunag:arXiv:2010} However, varying surface roughness is often
accompanied by varying crystallinity, and therefore the results do
not allow for a clear  distinction between the contributions
of the surface and the bulk.

In this paper we provide evidence from experiment, theory, and
simulation that a moderate amount of surface roughness has no
significant impact on the far-field optical properties of metallic
nanoparticles.  We interpret this as a kind of \textit{motional
narrowing}, where the surface plasmon averages over the random height
fluctuations of the metal surface, which leads to destructive
interference and an overall small net effect.  Our findings might be beneficial for the design of metamaterials, which rely on large ensembles of
practically identical particles, as well as devices based on the 
far-field properties of plasmonic nanoparticles.

\begin{figure}
\centerline{\includegraphics[width=0.6\columnwidth]{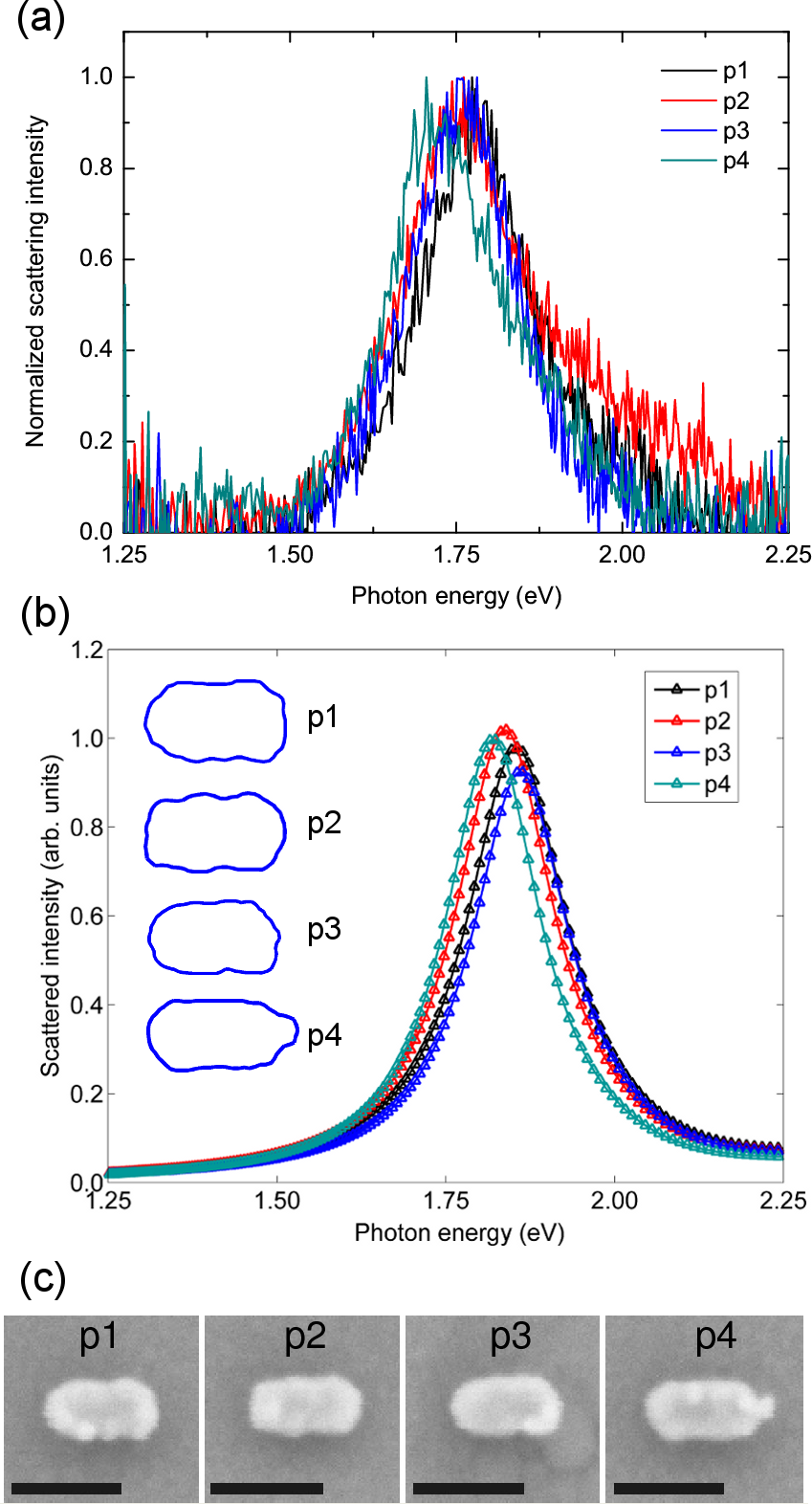}}
\caption{\label{fig:exp}(color online) (a) Scattering spectra of four individual gold nanorods excited with a polarization parallel to the particles long axis. Since the repeatability of the scattering intensity is only within $\sim 20\%$, the spectra are normalized to facilitate the spectral comparison.
(b) Spectra simulated within our BEM approach, using the gold dielectric function of Ref.~\onlinecite{johnson:72} and a refractive index of $n_b=1.65$ for the substrate, for the particle shapes extraced from the SEM images reported in panel (c). The particle height is $45\:{\rm nm}$.  The lengths of the scalebars in the lower panels are $100\:{\rm nm}$.}
\end{figure}

\textit{Experiment}.---%
We measured the scattering spectra of individual 
polycrystalline (rough) gold nanoparticles to probe the influence of 
unavoidable surface roughness on the plasmonic signature of 
nanoparticles of nominally identical shape. The particles were 
fabricated by electron beam lithography on an indium-tin oxyde (ITO) covered glass substrate.  The
substrate was coated with a polymer resist, which was then exposed and
chemically developed. Vacuum deposition of gold and a liftoff process  lead to
polycrystalline particles (crystallite size $\sim 20\:{\rm nm}$) of
designed  shapes,~\cite{Hohenau:MicroelEng:06} but with apparent surface
roughness, as shown in Fig.~\ref{fig:exp}(c). The particles  were 
probed in a dark field microscope, collecting the  scattered light by
a $40 \times$, $0.75$ numerical aperture objective  and analyzing it with
a spectrograph. 

Fig.~\ref{fig:exp}(a) shows exemplarily scattering spectra of four
individual $50\:{\rm nm}$ wide, $100\:{\rm nm}$ long, and $45\:{\rm nm}$
high gold nanorods.  Although the exact particle shapes differ due
to nanoscale roughness [Fig.~\ref{fig:exp}(c)],
the dipolar plasmon resonance positions and the spectral widths of the four
particles under consideration differ less than 20\% of the full width 
at half maximum.  Similar results were also observed for other nanoparticles shapes
(not shown).  From this we conclude that surface roughness  has a surprisingly small influence on the optical properties of plasmonic nanoparticles, at least in the regime where the
roughness does not lead to a noticeable change of the particle aspect
ratio.


\textit{Simulation}.---%
We additionally performed simulations based on the boundary element method (BEM),~\cite{garcia:02,garcia:10,hohenester.prb:05,hohenester.ieee:08} using a dielectric function representative for gold~\cite{johnson:72} and an effective refractive index of $n_b=1.65$ for the ITO covered glass substrate.
The particle shapes in the $(x,y)$-plane were extracted from the SEM images [Fig.~1(c)].  We use a particle height of 45 nm and round off the edges with a curvature radius of 5 nm.  The simulated scattering spectra are reported in Fig.~1(b).  In accordance to experiment, we find a surprisingly small influence of the roughness-related particle shape on the plasmon peak positions.  Also the absolute peak heights vary by less than 5 percent.  We note that there are small differences between experiment and theory, such as the weak shoulder around 2 eV, but the overall agreement regarding both peak positions and widths is striking.

The agreement between the measured and calculated widths of the plasmon peaks is quite remarkable since in our simulations we do not consider plasmon dephasing due to surface roughness scatterings, but only include electrodynamic decay channels.  From supplementary calculations we find that the absorption cross sections are about three times larger than the scattering cross sections.  Thus, ohmic losses (described through the imaginary part of the dielectric function) dominate over radiative damping as well as over dephasing losses due to surface roughness scatterings.  For smaller particle sizes, comparable to the electron mean free path of several nanometers, such dephasing will become increasingly important and could no longer be neglected in the simulations.

To inquire into the reasons for the almost negligible influence of surface roughness on the plasmon peak positions, in the following we use cylindrical nanorods as a showcase system, because of their extreme sensitivity to variations of the dielectric environment and shape.~\cite{becker:10}  We also employ the quasistatic approximation,~\cite{garcia:02,fuchs:75} which is justified for small particles and will allow us further below to introduce a perturbation analysis.  With respect to the retarded BEM simulations, the quasistatic approximation is expected to lead to a slight overestimation of the plasmon peak energy.

\begin{figure}
\centerline{\includegraphics[width=0.65\columnwidth]{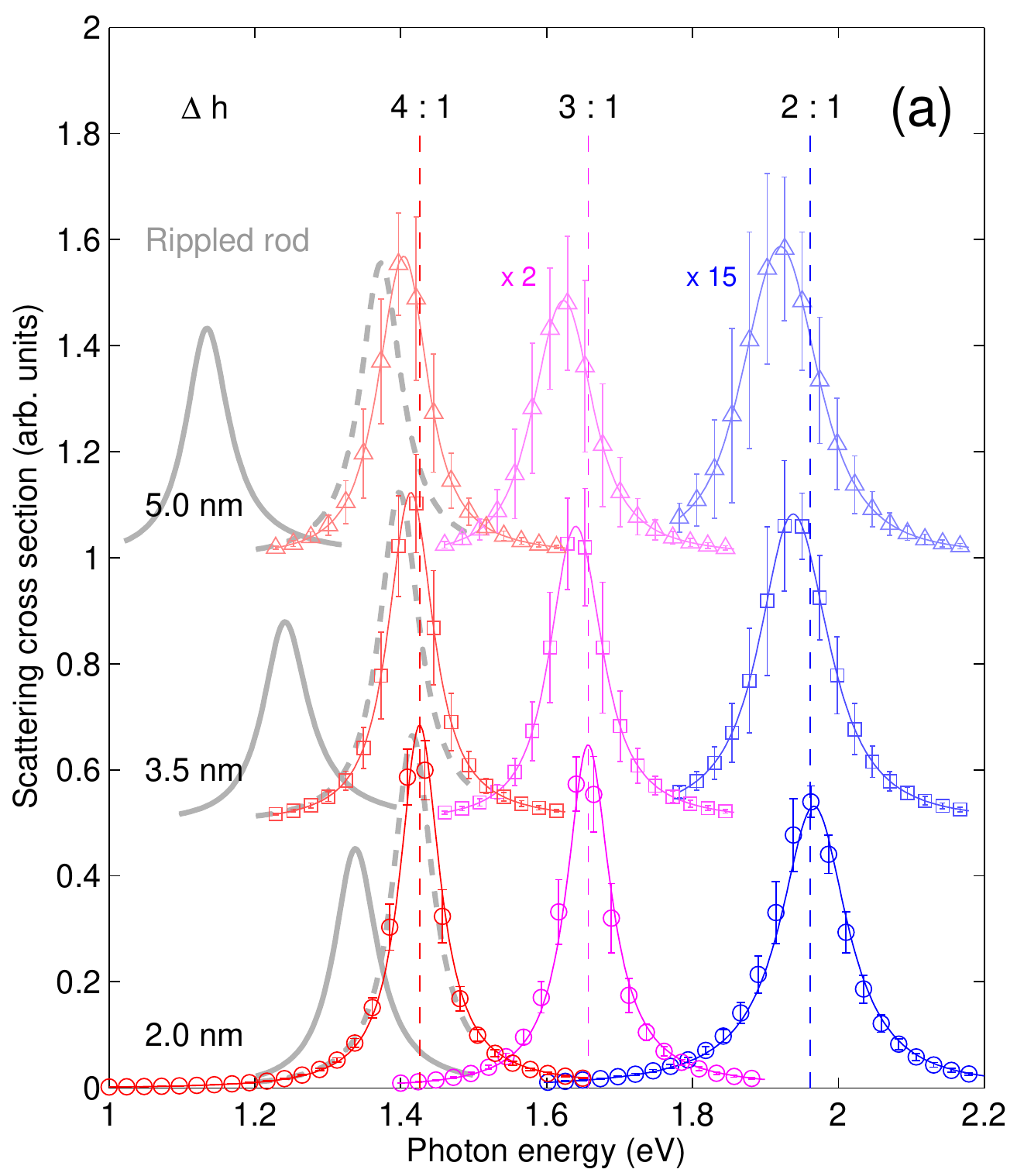}}
\centerline{\includegraphics[width=0.65\columnwidth]{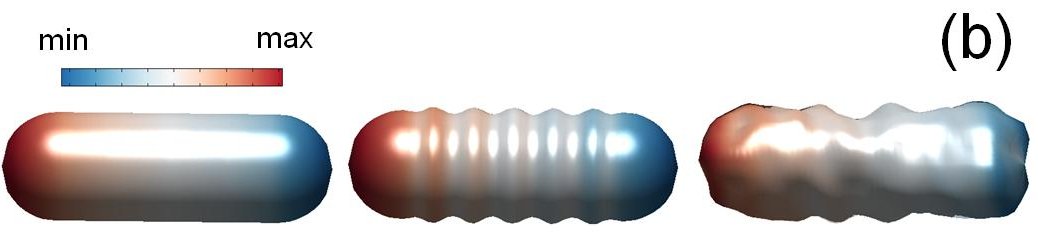}}
\caption{(color online) (a) Simulated spectra for nanorods with a diameter of 30 nm and for height-to-diameter ratios of $2:1$, $3:1$, and $4:1$.  In our simulations we use a dielectric function representative for gold~\cite{johnson:72} and a homogeneous dielectric background ($n_b=1.5$), and add stochastically surface roughness according to the prescription of Eq.~\eqref{eq:roughness} with $\sigma_h=3$ nm and for different $\Delta h$ values, which determine the height variations of surface roughness.  Each spectrum is averaged over 100 randomly generated nanoparticles.  The solid lines report the spectra of the smooth nanorods in the panel for $\Delta h=2$ nm, and the mean values of the spectra otherwise.  The gray lines report results for rippled rods, as discussed in the text.  The spectra for ratios $2:1$ and $3:1$ have been magnified by factors of 15 and 2, respectively.  (b) Surface charge distributions $\sigma_a$ for the optically active plasmon modes for the smooth (left), rippled (middle), and rough (right) nanorods. }\label{fig:spectrum}
\end{figure}

In our simulations we model surface roughness by adding stochastic height variations to the smooth surface of an ideal nanoparticle.  In two dimensions and for a box with periodic boundary conditions, height variations with a Gaussian autocorrelation can be obtained by attaching to all Fourier coefficients arbitrary phase factors $e^{i\phi_{\rm rnd}}$ viz.
\begin{equation}\label{eq:roughness}
  h(x,y)=\Delta h\,\Re e\left[\mathcal{F}^{-1}\left(
    e^{-\frac 12{\sigma_h^2}\left(k_x^2+k_y^2\right)+i\phi_{\rm rnd}}
  \right)-\frac 12\right]\,,
\end{equation}
where $\mathcal{F}^{-1}$ denotes the inverse Fourier transform and $\sigma_h^2$ is the variance of the height fluctuations.  We next map $h(x,y)$ to the nanoparticle surface and displace the vertices of the nanoparticle along the surface normal directions.~\cite{comment:map3d}  A typical realization of surface roughness is depicted in Fig.~\ref{fig:spectrum}(b) for a nanorod. 

Figure~\ref{fig:spectrum}(a) shows spectra for gold nanorods for different height-to-diameter ratios.  Throughout we set the rod diameter to 30 nm and use $\sigma_h=3$ nm.  For each spectrum we average over hundred random realizations of surface roughness and assume light excitation with a polarization along long axis of the rod.  One clearly observes that when increasing the amount of surface roughness, by choosing different values of $\Delta h$, the shape of the averaged spectra does not change (with exception of an overall small red shift).  Only the variance of the individual spectra increases with increasing $\Delta h$, as indicated by the errorbars.  Thus, our numerical simulations demonstrate that \textit{surface roughness has a surprisingly small influence on the optical properties of plasmonic nanoparticles}.

Things change considerably if we use the more regular height variations of a rippled rod, depicted in Fig.~\ref{fig:spectrum}, that were recently suggested as a viable model for surface roughness.~\cite{pecharroman:08} Here the spectra strongly shift to the red when $\Delta h$ is increased, as depicted by the gray lines on the left of Fig.~\ref{fig:spectrum}.  Thus, it appears that the irregular shape of the surface height fluctuations is responsible for the overall small shift of the plasmon peak position.  Indeed, when ripples are introduced not only along the symmetry axis of the nanorod but also around its circumference (here 6 ripples) the redshift becomes strongly reduced, as shown by the dashed gray lines.


\textit{Theory}.---%
We next develop a perturbation analyis which will help us to understand the effects of surface roughness more deeply.  Let us first recall the basic elements of the BEM approach.  For a metallic nanoparticle with dielectric function $\epsilon_m(\omega)$ embedded in a dielectric background with constant $\epsilon_b$, the solutions of the Poisson equation within the two media are given by the Green function $G(\bm r,\bm r')=1/|\bm r-\bm r'|$.  The electrostatic potential can be written in the ad-hoc form $\phi=\bigl< G,\sigma\bigr>+\phi_{\rm ext}$,~\cite{garcia:02} where $\sigma$ is a surface charge distribution located at the boundary $\partial V$ of the metallic nanoparticle, which has to be chosen such that the boundary conditions of Maxwell's equations are fulfilled, $\left<G,\sigma\right>=\int_{\partial V}G(\bm r,\bm s)\sigma(\bm s)\,d\bm s$ defines an inner product, and $\phi_{\rm ext}$ is the potential of the external perturbation.

Continuity of the normal component of the dielectric displacement at the metal-dielectric interface gives an expression 
\begin{equation}\label{eq:surfacecharge}
  \Lambda(\omega)\sigma
  +\left<\frac{\partial G}{\partial\hat n},\sigma\right>=-
  \frac{\partial\phi_{\rm ext}}{\partial \hat n}\,,\quad
  \Lambda=2\pi \frac{\epsilon_m+\epsilon_b}{\epsilon_m-\epsilon_b}
\end{equation}
that can be used for the calculation of the surface charge distribution.  Here ${\partial G}/{\partial\hat n}\equiv F$ and ${\partial\phi_{\rm ext}}/{\partial \hat n}\equiv\phi_{\rm ext}'$ denote the surface derivatives of the Green function and the external potential, respectively.  

We next define the right and left eigenvectors $\sigma_k$ and $\tilde\sigma_k$ of the surface derivative of the Green function through~\cite{fuchs:75,mayergoyz:07}
\begin{equation}
  \bigl<F,\sigma_k\bigr>=\lambda_k\,\sigma_k\,,\quad
  \bigl<\tilde\sigma_k,F\bigr>=\lambda_k\,\tilde\sigma_k\,,
\end{equation}
which form a biorthogonal set with $\left<\tilde\sigma_k,\sigma_{k'}\right> =\delta_{kk'}$.  The functions $\sigma_k$ can be interpreted as the surface plasmon \textit{eigenmodes}, and the response to any external perturbation can be decomposed into these modes viz.
\begin{equation}\label{eq:eigenmodes}
  \sigma=-\sum_k 
  \frac{\sigma_k} {\Lambda(\omega)+\lambda_k}\,
  {\bigl<\tilde\sigma_k,\phi_{\rm ext}'\bigr>}\,.
\end{equation}
Apparently, a given mode $k$ gives a noticeable contribution only if the coupling $\left<\tilde\sigma_k,\phi_{\rm ext}'\right>$ to the external potential (in our case plane wave illumination) is sufficiently strong and if the denominator becomes small.  The plasmon resonance condition translates to $\mbox{Re}\left[\Lambda(\omega)+ \lambda_k\right]=0$, when assuming that the imaginary part of $\epsilon$ has an only weak frequency dependence.  This is an extremely useful expression as it allows to separate the structural properties (described by $\lambda_k$) from the material properties (described by $\Lambda$).

We are now in the position to analyze the effects of a moderate surface roughness, which we model as a distortion of the surface $\partial V$ from its ideal shape.  In turn, the surface derivative of the Green function $F$ changes to $F+\delta F$, where $\delta F$ is expected to have the same random character as the surface fluctuations.~\cite{comment:curvature}  How does $\lambda_k$, which determines the peak positions of the plasmons, change in case of surface roughness?  When $\delta F$ is sufficiently small we can employ perturbation theory, in complete analogy to quantum mechanics.  We treat $F$ as the unperturbed part and $\delta F$ as the ``perturbation''.  Following Ref.~\onlinecite{messiah:65} we introduce for a given plasmon mode $a$ the projector
\begin{equation}
  \frac {Q_0}a\equiv\sum_{k\neq a}\frac{\sigma_k^0\tilde\sigma_k^0}%
  {\lambda_a^0-\lambda_k^0}\,,
\end{equation}
where the superscript $0$ indicates the eigenvalues and eigenfunctions for the ideal nanoparticle surface.  The corrections to $\lambda_a$ then become within lowest order perturbation theory
\begin{subequations}\label{eq:lambda}
\begin{eqnarray}
  \lambda_a^1 &=& \Bigl<\tilde\sigma_a^0,\,\delta F\sigma_a^0\Bigr>\\
  \lambda_a^2 &=& \Bigl<\tilde\sigma_a^0,\,
    \delta F\frac {Q_0} a\delta F\,\sigma_a^0\Bigr>\,.
\end{eqnarray}
\end{subequations}
From this we can draw an important conclusion for the dipolar modes, whose $\sigma_a^0$ distributions usually vary smoothly on the length scale of surface roughness.  Since Eq.~(\ref{eq:lambda}a) averages the random variations $\delta F$ over the unperturbed distributions $\sigma_a^0$, the positive and negative $\delta F$ values associated with the height variations of surface roughness will become effectively averaged out.  This results in a small $\lambda_a^1$ correction, and consequently \textit{surface roughness does not affect significantly the optically active plasmon modes}.

In a sense, this finding is similar to motional narrowing in semiconductor quantum wells,~\cite{maialle:93} where the motion of an optically excited electron-hole pair (exciton) is subject to the potential induced by the local monolayer fluctuation inherent to quantum wells.  When the exciton propagates through the well it ``averages'' over the fluctuations of the random potential, which results in a narrowing of the exciton lineshape. 

\begin{figure}
\centerline{\includegraphics[width=0.7\columnwidth]{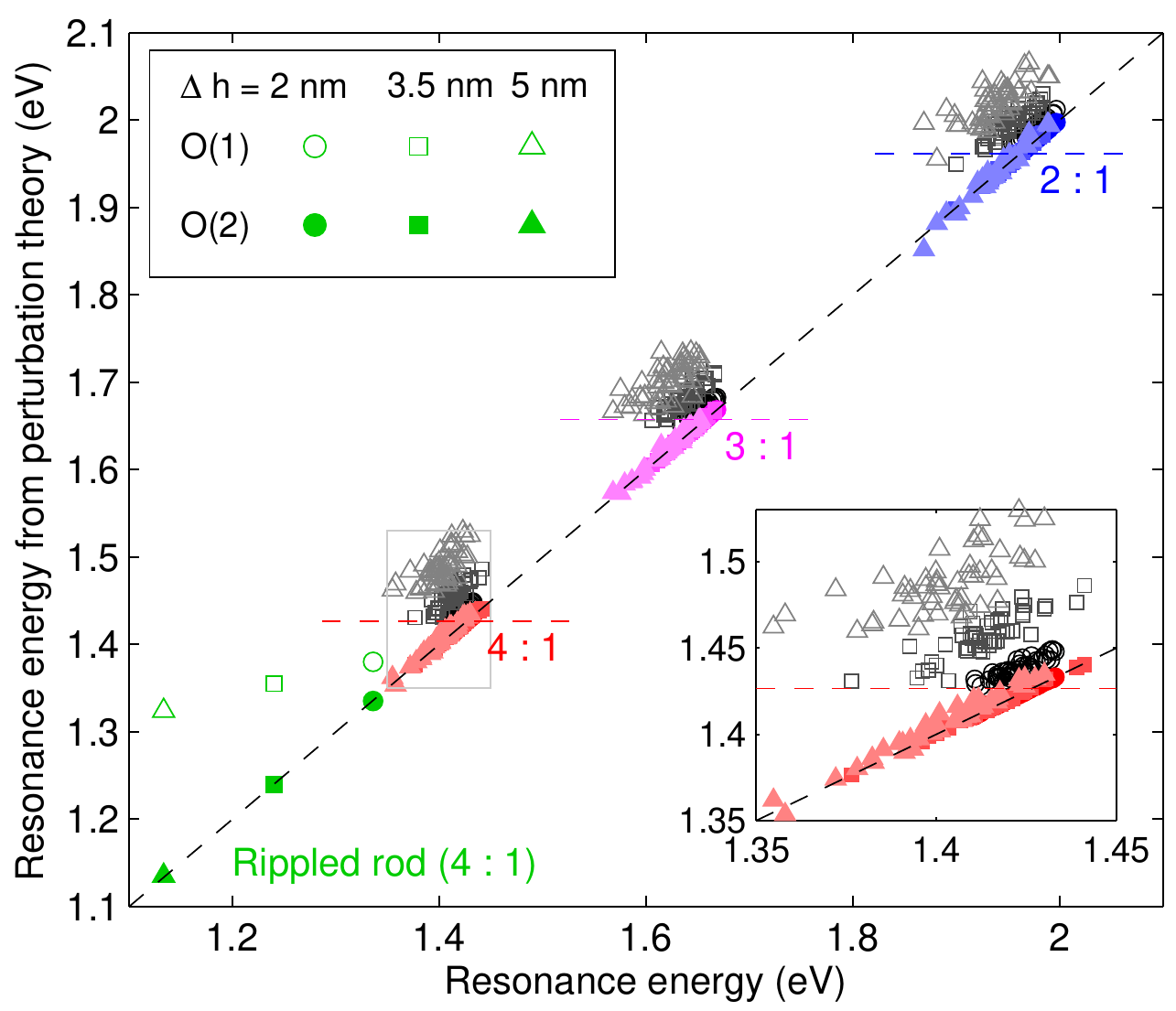}}
\caption{(color online) Comparison of full simulations and perturbation theory of Eq.~\eqref{eq:lambda}, for the positions of the resonance energies of the optically active surface plasmon modes with polarization along the long axis of the rod.  The open and filled symbols report results obtained within first and second order perturbation theory, respectively.  In the inset we show a blow-up for the nanorod with a height-to-diameter ratio of $4:1$.}\label{fig:perturbation}
\end{figure}

In Fig.~\ref{fig:perturbation} we compare the results of our full simulations with the predictions of Eq.~\eqref{eq:lambda}.  The open and full symbols show the true resonance energies (obtained from $\mbox{Re}\left[\Lambda(\omega)+ \lambda_k\right]=0$) and the perturbation results (obtained with $\lambda_a^0+\lambda_a^1$ and $\lambda_a^0+\lambda_a^1+ \lambda_a^2$, respectively).  Symbols on the diagonal correspond to the situation where perturbation theory and the full BEM simulations coincide.  In all cases the shift of the resonance with respect to the positions of the smooth nanorod (dashed horizontal lines) is small.  Quite generally, first order perturbation theory, which ignores any variations of the surface charge distribution ($\sigma_a=\sigma_a^0$), leads to a blue shift.  Only for second order perturbation theory, which includes modification of $\sigma_a$, the red shift of the plasmonic resonances is properly reproduced. 

As for the rippled rod, we observe that the first-order corrections $\lambda_a^1$ are small.  This is because the argument of motional narrowing in principle also applies here.  However, through the admixture of excited surface plasmon modes $\sigma^0$, described by the second-order correction $\lambda_a^2$, the surface plasmon can accomodate to the regular height variations of the rippled rod, and the plasmon peak position becomes strongly red-shifted.  No corresponding conclusions prevail for the stochastic height variations.


\textit{Summary}.---%
In summary, we have investigated the influence of surface roughness on the optical properties of plasmonic nanoparticles, and have found a surprisingly small effect.  Using a simulation and perturbation theory approach, we have been able to trace back our findings to a motional narrowing, where the plasmon averages over the random height fluctuations.  As no corresponding conclusions prevail for the near-field optical properties, our results are in accordance with the findings of ``hot spots'' in fluorescence or surface enhanced Raman scattering experiments.  

This work has been supported in part by the Austrian science fund FWF under project No. P21235--N20.


\begin{thebibliography}{26}
\expandafter\ifx\csname natexlab\endcsname\relax\def\natexlab#1{#1}\fi
\expandafter\ifx\csname bibnamefont\endcsname\relax
  \def\bibnamefont#1{#1}\fi
\expandafter\ifx\csname bibfnamefont\endcsname\relax
  \def\bibfnamefont#1{#1}\fi
\expandafter\ifx\csname citenamefont\endcsname\relax
  \def\citenamefont#1{#1}\fi
\expandafter\ifx\csname url\endcsname\relax
  \def\url#1{\texttt{#1}}\fi
\expandafter\ifx\csname urlprefix\endcsname\relax\def\urlprefix{URL }\fi
\providecommand{\bibinfo}[2]{#2}
\providecommand{\eprint}[2][]{\url{#2}}

\bibitem[{\citenamefont{Maier}(2007)}]{maier:07}
\bibinfo{author}{\bibfnamefont{S.~A.} \bibnamefont{Maier}},
  \emph{\bibinfo{title}{Plasmonics: Fundamentals and Applications}}
  (\bibinfo{publisher}{Springer}, \bibinfo{address}{Berlin},
  \bibinfo{year}{2007}).

\bibitem[{\citenamefont{Schuller et~al.}(2010)\citenamefont{Schuller, Barnard,
  Cai, Jun, White, and Brongersma}}]{Schuller:NatureMat:2010}
\bibinfo{author}{\bibfnamefont{J.~A.} \bibnamefont{Schuller}},
  \bibinfo{author}{\bibfnamefont{E.~S.} \bibnamefont{Barnard}},
  \bibinfo{author}{\bibfnamefont{W.}~\bibnamefont{Cai}},
  \bibinfo{author}{\bibfnamefont{Y.~C.} \bibnamefont{Jun}},
  \bibinfo{author}{\bibfnamefont{J.~S.} \bibnamefont{White}}, \bibnamefont{and}
  \bibinfo{author}{\bibfnamefont{M.~L.} \bibnamefont{Brongersma}},
  \bibinfo{journal}{Nature Materials} \textbf{\bibinfo{volume}{9}},
  \bibinfo{pages}{193} (\bibinfo{year}{2010}).

\bibitem[{\citenamefont{Anger et~al.}(2006)\citenamefont{Anger, Bharadwaj, and
  Novotny}}]{anger:06}
\bibinfo{author}{\bibfnamefont{P.}~\bibnamefont{Anger}},
  \bibinfo{author}{\bibfnamefont{P.}~\bibnamefont{Bharadwaj}},
  \bibnamefont{and} \bibinfo{author}{\bibfnamefont{L.}~\bibnamefont{Novotny}},
  \bibinfo{journal}{Phys. Rev. Lett.} \textbf{\bibinfo{volume}{96}},
  \bibinfo{pages}{113002} (\bibinfo{year}{2006}).

\bibitem[{\citenamefont{Kneipp et~al.}(2006)\citenamefont{Kneipp, Moskovits,
  and Kneipp}}]{Kneipp:Book}
\bibinfo{editor}{\bibfnamefont{K.}~\bibnamefont{Kneipp}},
  \bibinfo{editor}{\bibfnamefont{M.}~\bibnamefont{Moskovits}},
  \bibnamefont{and} \bibinfo{editor}{\bibfnamefont{H.}~\bibnamefont{Kneipp}},
  eds., \emph{\bibinfo{title}{Surface-Enhanced Raman Scattering}}, vol.
  \bibinfo{volume}{103} of \emph{\bibinfo{series}{{\em Topics in Applied
  Physics}}} (\bibinfo{publisher}{Springer}, \bibinfo{address}{Heidelberg, New
  York}, \bibinfo{year}{2006}).

\bibitem[{\citenamefont{Anker et~al.}(2008)\citenamefont{Anker, Hall, Lyandres,
  Shah, Zhao, and Duyne}}]{anker:08}
\bibinfo{author}{\bibfnamefont{J.~N.} \bibnamefont{Anker}},
  \bibinfo{author}{\bibfnamefont{W.~P.} \bibnamefont{Hall}},
  \bibinfo{author}{\bibfnamefont{O.}~\bibnamefont{Lyandres}},
  \bibinfo{author}{\bibfnamefont{N.~C.} \bibnamefont{Shah}},
  \bibinfo{author}{\bibfnamefont{J.}~\bibnamefont{Zhao}}, \bibnamefont{and}
  \bibinfo{author}{\bibfnamefont{R.~P.~V.} \bibnamefont{Duyne}},
  \bibinfo{journal}{Nature Mater.} \textbf{\bibinfo{volume}{7}},
  \bibinfo{pages}{442} (\bibinfo{year}{2008}).

\bibitem[{\citenamefont{Ferry et~al.}(2008)\citenamefont{Ferry, Sweatlock,
  Pacifici, and Atwater}}]{ferry:08}
\bibinfo{author}{\bibfnamefont{V.~E.} \bibnamefont{Ferry}},
  \bibinfo{author}{\bibfnamefont{L.~A.} \bibnamefont{Sweatlock}},
  \bibinfo{author}{\bibfnamefont{D.}~\bibnamefont{Pacifici}}, \bibnamefont{and}
  \bibinfo{author}{\bibfnamefont{H.~A.} \bibnamefont{Atwater}},
  \bibinfo{journal}{Nano Lett.} \textbf{\bibinfo{volume}{8}},
  \bibinfo{pages}{4391} (\bibinfo{year}{2008}).

\bibitem[{\citenamefont{Chang et~al.}(2006)\citenamefont{Chang, Sorensen,
  Hemmer, and Lukin}}]{Chang:PhysRevLett:06}
\bibinfo{author}{\bibfnamefont{D.~E.} \bibnamefont{Chang}},
  \bibinfo{author}{\bibfnamefont{A.~S.} \bibnamefont{Sorensen}},
  \bibinfo{author}{\bibfnamefont{P.~R.} \bibnamefont{Hemmer}},
  \bibnamefont{and} \bibinfo{author}{\bibfnamefont{M.~D.} \bibnamefont{Lukin}},
  \bibinfo{journal}{Phys. Rev. Lett.} \textbf{\bibinfo{volume}{97}},
  \bibinfo{pages}{053002} (\bibinfo{year}{2006}).

\bibitem[{\citenamefont{Pendry et~al.}(2006)\citenamefont{Pendry, Schurig, and
  Smith}}]{pendry:06}
\bibinfo{author}{\bibfnamefont{J.~B.} \bibnamefont{Pendry}},
  \bibinfo{author}{\bibfnamefont{D.}~\bibnamefont{Schurig}}, \bibnamefont{and}
  \bibinfo{author}{\bibfnamefont{D.~R.} \bibnamefont{Smith}},
  \bibinfo{journal}{Science} \textbf{\bibinfo{volume}{312}},
  \bibinfo{pages}{1780} (\bibinfo{year}{2006}).

\bibitem[{\citenamefont{Rodriguez-Fernandez
  et~al.}(2009)\citenamefont{Rodriguez-Fernandez, Funston, Perez-Juste,
  Alvarez-Puebla, Liz-Marzan, and Mulvaney}}]{Marzan:CPPC:2009}
\bibinfo{author}{\bibfnamefont{J.}~\bibnamefont{Rodriguez-Fernandez}},
  \bibinfo{author}{\bibfnamefont{A.~M.} \bibnamefont{Funston}},
  \bibinfo{author}{\bibfnamefont{J.}~\bibnamefont{Perez-Juste}},
  \bibinfo{author}{\bibfnamefont{R.~A.} \bibnamefont{Alvarez-Puebla}},
  \bibinfo{author}{\bibfnamefont{L.~M.} \bibnamefont{Liz-Marzan}},
  \bibnamefont{and} \bibinfo{author}{\bibfnamefont{P.}~\bibnamefont{Mulvaney}},
  \bibinfo{journal}{Phys. Chem. Chem. Phys.} \textbf{\bibinfo{volume}{11}},
  \bibinfo{pages}{5909} (\bibinfo{year}{2009}).

\bibitem[{\citenamefont{Chen et~al.}(2009)\citenamefont{Chen, Drachev,
  Borneman, Kildishev, and Shalaev}}]{Chen:NanoLetters:09}
\bibinfo{author}{\bibfnamefont{K.-P.} \bibnamefont{Chen}},
  \bibinfo{author}{\bibfnamefont{V.~P.} \bibnamefont{Drachev}},
  \bibinfo{author}{\bibfnamefont{J.~D.} \bibnamefont{Borneman}},
  \bibinfo{author}{\bibfnamefont{A.~V.} \bibnamefont{Kildishev}},
  \bibnamefont{and} \bibinfo{author}{\bibfnamefont{V.~M.}
  \bibnamefont{Shalaev}}, \bibinfo{journal}{Nano Letters} p.
  \bibinfo{pages}{916} (\bibinfo{year}{2009}).

\bibitem[{\citenamefont{Barnes}(2009)}]{Barnes:JOptA:09}
\bibinfo{author}{\bibfnamefont{W.~L.} \bibnamefont{Barnes}},
  \bibinfo{journal}{J. Opt. A.} \textbf{\bibinfo{volume}{11}},
  \bibinfo{pages}{114002} (\bibinfo{year}{2009}).

\bibitem[{\citenamefont{Huang et~al.}(2010)\citenamefont{Huang, Callegari,
  Geisler, Br\"uning, Kern, Prangsma, Weinmann, Kamp, Forchel, Biagioni
  et~al.}}]{Hunag:arXiv:2010}
\bibinfo{author}{\bibfnamefont{J.~S.} \bibnamefont{Huang}},
  \bibinfo{author}{\bibfnamefont{V.}~\bibnamefont{Callegari}},
  \bibinfo{author}{\bibfnamefont{P.}~\bibnamefont{Geisler}},
  \bibinfo{author}{\bibfnamefont{C.}~\bibnamefont{Br\"uning}},
  \bibinfo{author}{\bibfnamefont{J.}~\bibnamefont{Kern}},
  \bibinfo{author}{\bibfnamefont{J.}~\bibnamefont{Prangsma}},
  \bibinfo{author}{\bibfnamefont{P.}~\bibnamefont{Weinmann}},
  \bibinfo{author}{\bibfnamefont{M.}~\bibnamefont{Kamp}},
  \bibinfo{author}{\bibfnamefont{A.}~\bibnamefont{Forchel}},
  \bibinfo{author}{\bibfnamefont{P.}~\bibnamefont{Biagioni}},
  \bibnamefont{et~al.}, \bibinfo{journal}{arXiv:1004.1961}
  (\bibinfo{year}{2010}).

\bibitem[{\citenamefont{Hohenau et~al.}(2006)\citenamefont{Hohenau, Ditlbacher,
  Lamprecht, Krenn, Leitner, and Aussenegg}}]{Hohenau:MicroelEng:06}
\bibinfo{author}{\bibfnamefont{A.}~\bibnamefont{Hohenau}},
  \bibinfo{author}{\bibfnamefont{H.}~\bibnamefont{Ditlbacher}},
  \bibinfo{author}{\bibfnamefont{B.}~\bibnamefont{Lamprecht}},
  \bibinfo{author}{\bibfnamefont{J.~R.} \bibnamefont{Krenn}},
  \bibinfo{author}{\bibfnamefont{A.}~\bibnamefont{Leitner}}, \bibnamefont{and}
  \bibinfo{author}{\bibfnamefont{F.~R.} \bibnamefont{Aussenegg}},
  \bibinfo{journal}{Microelec. Engin.} \textbf{\bibinfo{volume}{83}},
  \bibinfo{pages}{1464} (\bibinfo{year}{2006}).

\bibitem[{\citenamefont{Johnson and Christy}(1972)}]{johnson:72}
\bibinfo{author}{\bibfnamefont{P.~B.} \bibnamefont{Johnson}} \bibnamefont{and}
  \bibinfo{author}{\bibfnamefont{R.~W.} \bibnamefont{Christy}},
  \bibinfo{journal}{Phys. Rev. B} \textbf{\bibinfo{volume}{6}},
  \bibinfo{pages}{4370} (\bibinfo{year}{1972}).

\bibitem[{\citenamefont{Garci{a de Abajo} and Howie}(2002)}]{garcia:02}
\bibinfo{author}{\bibfnamefont{F.~J.} \bibnamefont{Garci{a de Abajo}}}
  \bibnamefont{and} \bibinfo{author}{\bibfnamefont{A.}~\bibnamefont{Howie}},
  \bibinfo{journal}{Phys. Rev. B} \textbf{\bibinfo{volume}{65}},
  \bibinfo{pages}{115418} (\bibinfo{year}{2002}).

\bibitem[{\citenamefont{Hohenester and Krenn}(2005)}]{hohenester.prb:05}
\bibinfo{author}{\bibfnamefont{U.}~\bibnamefont{Hohenester}} \bibnamefont{and}
  \bibinfo{author}{\bibfnamefont{J.~R.} \bibnamefont{Krenn}},
  \bibinfo{journal}{Phys. Rev. B} \textbf{\bibinfo{volume}{72}},
  \bibinfo{pages}{195429} (\bibinfo{year}{2005}).

\bibitem[{\citenamefont{Hohenester and Tr{\"u}gler}(2008)}]{hohenester.ieee:08}
\bibinfo{author}{\bibfnamefont{U.}~\bibnamefont{Hohenester}} \bibnamefont{and}
  \bibinfo{author}{\bibfnamefont{A.}~\bibnamefont{Tr{\"u}gler}},
  \bibinfo{journal}{IEEE J. of Selected Topics in Quantum Electronics}
  \textbf{\bibinfo{volume}{14}}, \bibinfo{pages}{1430} (\bibinfo{year}{2008}).

\bibitem[{\citenamefont{de~Abajo}(2010)}]{garcia:10}
\bibinfo{author}{\bibfnamefont{F.~J.~G.} \bibnamefont{de~Abajo}},
  \bibinfo{journal}{Rev. Mod. Phys.} \textbf{\bibinfo{volume}{82}},
  \bibinfo{pages}{209} (\bibinfo{year}{2010}).

\bibitem[{\citenamefont{Becker et~al.}(2010)\citenamefont{Becker, Tr{\"u}gler,
  Jakab, Hohenester, and S{\"o}nnichsen}}]{becker:10}
\bibinfo{author}{\bibfnamefont{J.}~\bibnamefont{Becker}},
  \bibinfo{author}{\bibfnamefont{A.}~\bibnamefont{Tr{\"u}gler}},
  \bibinfo{author}{\bibfnamefont{A.}~\bibnamefont{Jakab}},
  \bibinfo{author}{\bibfnamefont{U.}~\bibnamefont{Hohenester}},
  \bibnamefont{and}
  \bibinfo{author}{\bibfnamefont{C.}~\bibnamefont{S{\"o}nnichsen}},
  \bibinfo{journal}{Plasmonics} \textbf{\bibinfo{volume}{5}},
  \bibinfo{pages}{161} (\bibinfo{year}{2010}).

\bibitem[{\citenamefont{Pecharrom\'an et~al.}(2008)\citenamefont{Pecharrom\'an,
  P\'erez-Juste, Mata-Osoro, Liz-Marz\'an, and Mulvaney}}]{pecharroman:08}
\bibinfo{author}{\bibfnamefont{C.}~\bibnamefont{Pecharrom\'an}},
  \bibinfo{author}{\bibfnamefont{J.}~\bibnamefont{P\'erez-Juste}},
  \bibinfo{author}{\bibfnamefont{G.}~\bibnamefont{Mata-Osoro}},
  \bibinfo{author}{\bibfnamefont{L.~M.} \bibnamefont{Liz-Marz\'an}},
  \bibnamefont{and} \bibinfo{author}{\bibfnamefont{P.}~\bibnamefont{Mulvaney}},
  \bibinfo{journal}{Phys. Rev. B} \textbf{\bibinfo{volume}{77}},
  \bibinfo{pages}{035418} (\bibinfo{year}{2008}).

\bibitem[{\citenamefont{Fuchs}(1975)}]{fuchs:75}
\bibinfo{author}{\bibfnamefont{R.}~\bibnamefont{Fuchs}},
  \bibinfo{journal}{Phys. Rev. B} \textbf{\bibinfo{volume}{11}},
  \bibinfo{pages}{1732} (\bibinfo{year}{1975}).

\bibitem[{com({\natexlab{a}})}]{comment:map3d}
\bibinfo{note}{Mapping can be faciliated by computing $h(x,y,z)$ in three
  dimensions, with an expression similar to Eq.~\eqref{eq:roughness}, and
  interpolating the stochastic height variations $h$ to the nanoparticle
  surface.}

\bibitem[{\citenamefont{Mayergoyz et~al.}(2007)\citenamefont{Mayergoyz, Zhang,
  and Miano}}]{mayergoyz:07}
\bibinfo{author}{\bibfnamefont{I.~D.} \bibnamefont{Mayergoyz}},
  \bibinfo{author}{\bibfnamefont{Z.}~\bibnamefont{Zhang}}, \bibnamefont{and}
  \bibinfo{author}{\bibfnamefont{G.}~\bibnamefont{Miano}},
  \bibinfo{journal}{Phys. Rev. Lett.} \textbf{\bibinfo{volume}{98}},
  \bibinfo{pages}{147401} (\bibinfo{year}{2007}).

\bibitem[{com({\natexlab{b}})}]{comment:curvature}
\bibinfo{note}{This is because $F(\bm s,\bm s')$ is governed for $\bm
  s\approx\bm s'$, where deviations of the ideal shape are most significant,
  directly by the surface curvature.}

\bibitem[{\citenamefont{Messiah}(1965)}]{messiah:65}
\bibinfo{author}{\bibfnamefont{A.}~\bibnamefont{Messiah}},
  \emph{\bibinfo{title}{Quantum Mechanics}} (\bibinfo{publisher}{North-Holland,
  Amsterdam}, \bibinfo{year}{1965}).

\bibitem[{\citenamefont{Maialle et~al.}(1993)\citenamefont{Maialle, {de Andrada
  e Silva}, and Sham}}]{maialle:93}
\bibinfo{author}{\bibfnamefont{M.~Z.} \bibnamefont{Maialle}},
  \bibinfo{author}{\bibfnamefont{E.~A.} \bibnamefont{{de Andrada e Silva}}},
  \bibnamefont{and} \bibinfo{author}{\bibfnamefont{L.~J.} \bibnamefont{Sham}},
  \bibinfo{journal}{Phys. Rev. B} \textbf{\bibinfo{volume}{47}},
  \bibinfo{pages}{15 776} (\bibinfo{year}{1993}).

\end{thebibliography}
\end{document}